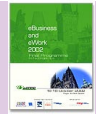

# An Agent Framework for Dynamic Agent Retraining: Agent Academy

P. Mitkas[1], A. Symeonidis[1], D. Kechagias[1], I. Athanasiadis[1],
G. Laleci[2], G. Kurt[2], Y. Kabak[2], A. Acar[2], A. Dogac[2]

[1] Intelligent Systems and Software Engineering Laboratory, ITI, CERTH, 1st km Thermi – Panorama Road,
P.O. Box 361, GR 570 01 Thermi Thessaloniki, Greece Tel: +30 310 996349 mitkas@vergina.eng.auth.gr
[2] Software Research and Development Center, Middle East Technical University, Inonu Bulvari,
METU(ODTU) Campus, 06500, Ankara, Turkey Tel:+90 312 2105598 asuman@srdc.metu.edu.tr

**Abstract.** Agent Academy (AA) aims to develop a multi-agent society that can train new agents for specific or general tasks, while constantly retraining existing agents in a recursive mode. The system is based on collecting information both from the environment and the behaviors of the acting agents and their related successes/failures to generate a body of data, stored in the Agent Use Repository, which is mined by the Data Miner module, in order to generate useful knowledge about the application domain. Knowledge extracted by the Data Miner is used by the Agent Training Module as to train new agents or to enhance the behavior of agents already running.

In this paper the Agent Academy framework is introduced, and its overall architecture and functionality are presented. Training issues as well as agent ontologies are discussed. Finally, a scenario, which aims to provide environmental alerts to both individuals and public authorities, is described an AA-based use case.

## 1. Introduction

In the recent years agent technology has found many interesting applications in e-commerce, decision support systems and Internet applications. Several agent system building tools and frameworks have been developed. The feature lacking so far, though, is the ability to store, analyze and learn from the past activities of the agent society as a whole in order to dynamically re-train individual agents at run-time. In this way functionally unrelated agents within the society may benefit from each other's findings and be able to collectively exploit the shared knowledge base thereby increasing the effectiveness of the system. We claim that such a society-wide evaluation and constant learning approach is essential for describing complex and ultimately useful scenarios involving agents with different roles. In this paper, we provide the description of a framework capable of the said functionality which is being developed within the scope of IST-2000-31050 Agent Academy Project. The features of the proposed system are explained over a motivating scenario.

Agents have proven to be particularly useful in a series of business applications. In particular, agents have facilitated the procedure of buying and selling of goods and services in electronic marketplaces, the handling of workflows and have assisted in personalization by managing user profiles or by tackling production planning. As the benefits from using agent societies in such applications become clear, so does the need for development of high-level agent system-building tools.

There are numerous agent development platforms with various degrees of abstraction and completeness ranging from bare bones API's to full-fledged but less flexible building packages. These frameworks all have their strong points and lacking features, the most important factors being compliance to a specification (e.g. FIPA), support for the mobility of the agents from host to host, ability to support AI oriented languages (e.g. Lisp, JESS, etc.), support for lightweight agents, support for advanced planning-scheduling and combinations thereof. The feature lacking so far however is the ability to get feedback from an agent society, which will be analyzed in order to dynamically improve agents' intelligence at run-time. The Agent Academy Project aims to fill in this void.

## 2. The Agent Academy Framework

Agent Academy aims to develop a multi-agent system, which can train new agents for specific or general tasks and re-train its own agents in a recursive mode. The main components of the system are given in Figure 1. The system is based on collecting information from both the environment and the behaviors of the acting agents and their related successes/failures to generate a body of data. Then this body of data is mined through the Data Miner in order to discover the important relationships between the behaviors of the agents and their environment. These relationships discovered by the Data Miner are used by the training module, in order to train new agents or to retrain (i.e. enhance) the agents already running.

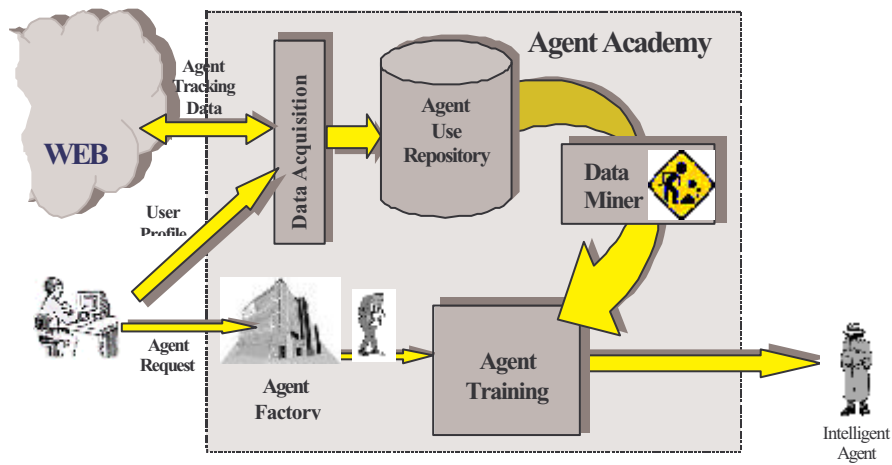

**Figure 1 Overall Architecture of the Agent Academy System**

The four main components of the Agent Academy framework are described below.

**Agent Factory (AF):** The Agent Factory is assigned with the creation of new (untrained) agents. User defines through an interface new agent's parameters, such as functional characteristics, abilities to access entities or devices, etc. The AF initializes all agents of the community according to user requests.

**Agent Use Repository (AUR):** The Agent Use Repository is a database holding all necessary data for a specific application. This body of data consists of agent behaviors, experience and decisions, as well as of information on the working environment and its parameters. AUR is constantly updated and appended with information via the *Data Acquisition Module*, which monitors and posts data from the operating agents.

**Data Mining Module (DM):** Through the data mining module a user can take advantage of data mining techniques after applying them to the body of application data. Results of the data mining process (in the form of decision trees, association rules and/or artificial neural networks) are posted to the ATM in FIPA-SL format, to be transplanted into active agents.

**Agent Training Module (ATM):** The Agent Training Module is charged with the task of embedding behaviors and beliefs to working agents. In this manner it is the tool for initially training new agents and for retraining AA produced active agents.

## 3. The Overall Architecture of the System

We decided to implement the AA framework using the Java Agent Development Environment (JADE) [1], which is a FIPA compliant Agent Development Toolkit. Another important step of the design concerning an agent based infrastructure is the setting of an Agent Communication Language, to allow the community agents to exchange information and services with one another, for negotiating, reasoning or co-operation. In the Agent Academy Framework we use FIPA-ACL (FIPA Agent Communication Language). In order to be fully FIPA compliant, the information in the Agent Communication Language message should be represented as a consistent content expression with one of the content languages in the FIPA Content Language Library [3], which are FIPA-SL (FIPA-Semantic Language), FIPA-RDF (FIPA-Resource Description Framework), FIPA-KIF (FIPA-Knowledge Interchange Format), and FIPA-CCL (FIPA-Constraint Choice Language). Agent Academy uses FIPA-SL [5], which is supported by JADE, as a content language to represent the information in ACL messages.

Another important issue about agent communication is the Ontology concept. The model of agent communication in FIPA is based on the assumption that two agents, who wish to converse, share a common ontology for the domain of discourse. It ensures that the agents ascribe the same meaning to the symbols used in the message. Java Agent Development Environment (JADE) provides a platform for easy exchange and parsing of these messages into internal representations of the agent by supporting ontology definitions consistent with the content language of the messages. Ontologies are defined as Java objects representing the internal structure of the message, and provide methods to retrieve the necessary parts of the message content. When an agent receives a Semantic Language (SL) message, which is in string format, it is parsed in to Java objects, and then checked with the previously defined ontologies. Then the agent extracts the components of the message by using the methods of the ontology. When sending a message, the agent constructs and initiates the necessary java objects and with the help of the content language support of JADE, the Semantic Language (SL) message in string format is constructed automatically.

In the following, the internal structure of the Agent Academy Framework is presented with the message structures in SL between the components. The scenario begins with a request to the Agent Factory (AF) for creating an agent. AF creates an untrained agent (UA) with the preliminary abilities to communicate with other agents, adding new behaviors to itself, being mobile if necessary. Then it sends a message to the Agent Training Module (ATM) giving the name and the type of the agent as shown in Table 1.

**Table 1 Agent Factory (AF)-Agent Training Module (ATM) interaction in SL format**

```
(agentsToBeTrained
     (agents (set (agent :name agent1
                         :type locationAgent))))
```

Upon receiving the message from Agent Factory, Agent Training Module discovers the necessary agent behaviors, which comes from the preferences of the user who requests the agent. ATM informs the Untrained Agent about these behaviors, which has the ability to load itself with new behaviors, with the message format presented in Table 2.

**Table 2 Agent Training Module - Untrained Agent interaction in Semantic Language (SL) Format**

```
(loadClass
     (behaviors (set  (behavior :classname Class1)
                      (behavior :classname Class2))))
```

In the Agent Academy Framework, the Untrained Agent that is trained by the Agent Training Module gains the ability to execute rules and perform reactive actions compatible with scenario-specific roles, as a result of these rules. While the agent performs its goals throughout its lifecycle, Data Acquisition Module collects statistical data, about the agent's behavior, successes or failures and updates the Agent Use Repository accordingly (e.g.

some rules are retracted, some new ones are added). Data Miner periodically imports data from the Agent Use Repository in order to generate associations, categorizations and sequences by using well-known Data Mining techniques, and outputs Decision Trees, Association Rules or Neural networks. These outputs are passed to the Agent Training Module to be taught to the agents that are executing in their environment. To pass these data structures, we defined SL representation formats for each of them.

The Agent Training Module re-inserts these enhanced rules to the both newly created untrained agents and to the existing agents to train them.

To be able to execute the rules and perform reactive actions accordingly, a rule engine is necessary. In our framework we are using Java Expert Systems Shell (JESS) [2] as the rule engine. The beliefs of the agents are represented as JESS rules, and when a new fact coming from the environment is asserted to the agent, these JESS rules are executed automatically and the action that is defined to be the result of this condition is executed. The agents created by Agent Factory gain the ability to execute JESS rules after an initial training by the Agent Training Module. The ATM, after having received the refined rules from the Data Mining Module as decision trees, neural networks, association rules etc, converts them into to JESS rules and sends them to the agents to be trained, in the message format presented in Table 3.

**Table 3 Semantic Language (SL) representation for the message format between ATM and Agent**

```
(addRule (jessRules (set   (jessRule :rule "(defrule rule_6 (and (ozone normal)) => (store ALARM_TYPE 3))")
                          (jessRule :rule "(defrule rule_5 (and (NO_2NO_3 normal)) => (store ALARM_TYPE 2))"))))
```

Operating with the new rules sent by the Agent Training Module (ATM), behavior of the agents would be enhanced in an iterative and recursive manner.

In the Agent Academy Framework, different scenarios can be implemented with different domain specific ontologies. For each ontology, there will be different databases in the Agent Use Repository. The field names of the databases correspond to the ontology term that the agent is internally using. Therefore the Data Miner needs to know on which ontology term to data mine. For example, in the O3RTAA scenario (described in Section 4), the ontology terms are pressure, ozone, nitrogen, alarm type and location. The attributes and the selection conditions that the Data Miner (DM) will consider and the attributes that the Agent uses should match syntactically.

One solution for the ontology problem is to define the ontology among the agents, which should be common as well as determined, prior to conducting the training protocol, while the agents are being created. Then the exchanged messages carrying the artificial neural network, the decision trees and other structures should also be constructed accordingly, agreeing on the common ontology of the scenario. The input and output of the data structures in the messages exchanged between the Artificial Neural Network and the Data Miner Module should agree with the terms of the ontology. Since both of the communicating agents use the same predefined ontology, the agents ascribe the same meaning to the symbols used in the message.

Another solution is to enable the agents to manage explicit, declaratively represented different ontologies. An ontology service for a community of agents can be specified for this purpose. A dedicated agent, called Ontology Agent, can provide this service. The role of the Ontology Agent can be summarized as follows [4]:

- ?? discovery of public ontologies in order to access them,
- ?? maintain a set of public ontologies (i.e. register with the Directory Facilitator of FIPA, upload, download, and modify),
- ?? translate expressions between different ontologies and/or different content languages,
- ?? respond to query for relationships between terms or between ontologies, and,
- ?? facilitate the identification of a shared ontology for communication between two agents.

With the help of the Ontology Agent, agents do not have to agree on a predefined ontology, but mappings between different ontologies can be provided by the Ontology Agent. Consider an agent getting the message that contains the term "pressure" and can not understand it since it is using the DomainSpecificOntology "*O3RTAATurkish*", whereas the sender agent is using the DomainSpecificOntology "*O3RTAAEnglish*". The agent can contact the Ontology Agent, asking the meaning of "pressure" in "*O3RTAATurkish*". The agent can achieve this by sending a Semantic Language (SL) message to the Ontology agent in the following format:

> (ontologyQuery (map :MessageOntology O3RTAAEnglish :MyOntology O3RTAATurkish :term pressure))

Then the ontology agent should inform the agent with the following SL message:

> (Mapping (From :term "pressure") (To :term "basinc"))

As a summary, the Agent Academy platform creates, trains and re-trains its society of agents in order to manage any given scenario. Such a scenario, which is the motivating application for the development, is the O3RTAA system, which uses agents for the collection and redistribution of environmental data.

## 4. The O3RTAA Scenario

The O3RTAA system aims to provide environmental alerts to both individuals (asthma patients, allergic people etc.) and public authorities (hospitals, civil protection etc.) depending on the levels of certain atmospheric variables, which are collected by the Mediterranean Centre for Environmental Studies (CEAM). The system is supposed to monitor the incoming data from the environmental sensors and make predictions as to the effects a certain change may cause in the environment. Some changes will lead to certain kind of alerts, which then need to be communicated effectively to parties interested in a particular alert.

The general structure of the O3RTAA system is given in Figure 2. The agents in this scenario are responsible for diagnosis, prediction and distribution. Each of these agent subsystems work as follows:

**Diagnosis Component:** This component will be in charge of analyzing the raw data coming in from the sensor data loggers. The main responsibility of the diagnosis agents is to screen this raw data and to forward only the significant changes while discarding faulty and redundant information. The human users may well input to the system through this component. And also, all the sensed information should be logged for future learning capability.

**Prediction Component:** When the diagnosis component decides that some change or fluctuation in the monitored parameters is significant, this change is forwarded to the prediction component. The predictor uses rule-bases and knowledge acquired from the historical data up to that point and also the current sensor information retrieved from the Diagnosis Component to determine what kind of hazard (i.e. alert) the change may be an indication of.

**Distribution Component:** When an alarm is raised by the prediction component, this event is sent to the distribution component which decides who will be affected by the alarm raised based on the profiles of the users and their current locations. Once a list of users to be informed is generated the alarm is sent to each user over the communication channel preferred by the user (e-mail, html, voice, SMS etc.). An important thing to consider here is that the following: According to the profile information and the urgency of the alarm, the distribution component decides on sending or delaying an alarm. As an example, a human authority may indicate in his profile that he wishes to receive the alarm through email and during office hours. However if an alarm is important, this component can decide to send an alarm at 4 A.M by a SMS (Short Message Service) to his phone number, and the email message later. Obviously if this decision is incorrect, this agent will be working inefficiently, but the importance of the alarm is decided by the Diagnosis component and the failure in this case is of the Diagnosis Component so the diagnosis component have have to be retrained through user feed back.

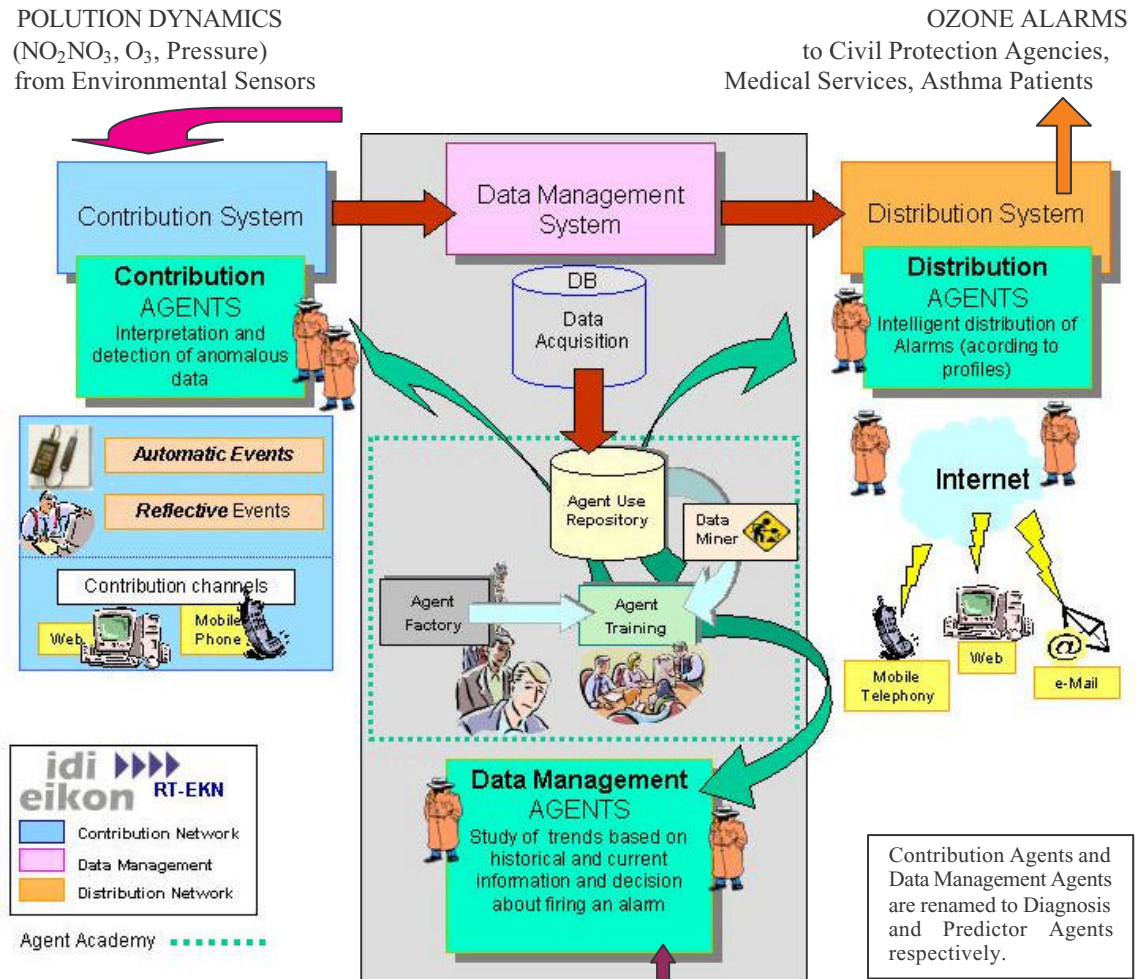

Automatic Network for Ambient Surveillance and Control,
once an hour checkings are provided from 25 Stations in Valencia region
**Figure 2 Overall View of the O3RTAA Scenario**

## 5. Realization of the O3RTAA scenario through the Agent Academy framework

How the Agent Academy framework handles the O3RTAA scenario can be explained through the processing steps when a new change in the environment occurs as shown in Figure 3.

Sensors collecting information on the environmental changes alert the Diagnosis Agent (1) when a change occurs in the parameters of the environment (i.e. $O_3$, $NO_2$, pressure, etc). The Diagnosis Agent, after validating the signal and deciding that it is significant, records the location information of the sensor and informs the Predictor Agent (2).

The Predictor Agent in the system, after obtaining the necessary information from the Diagnosis Agent, decides whether an alarm is to be raised and if so, what type of alarm is necessary and gives this to the Distributor agent (13) together with the location information to be distributed to the users.

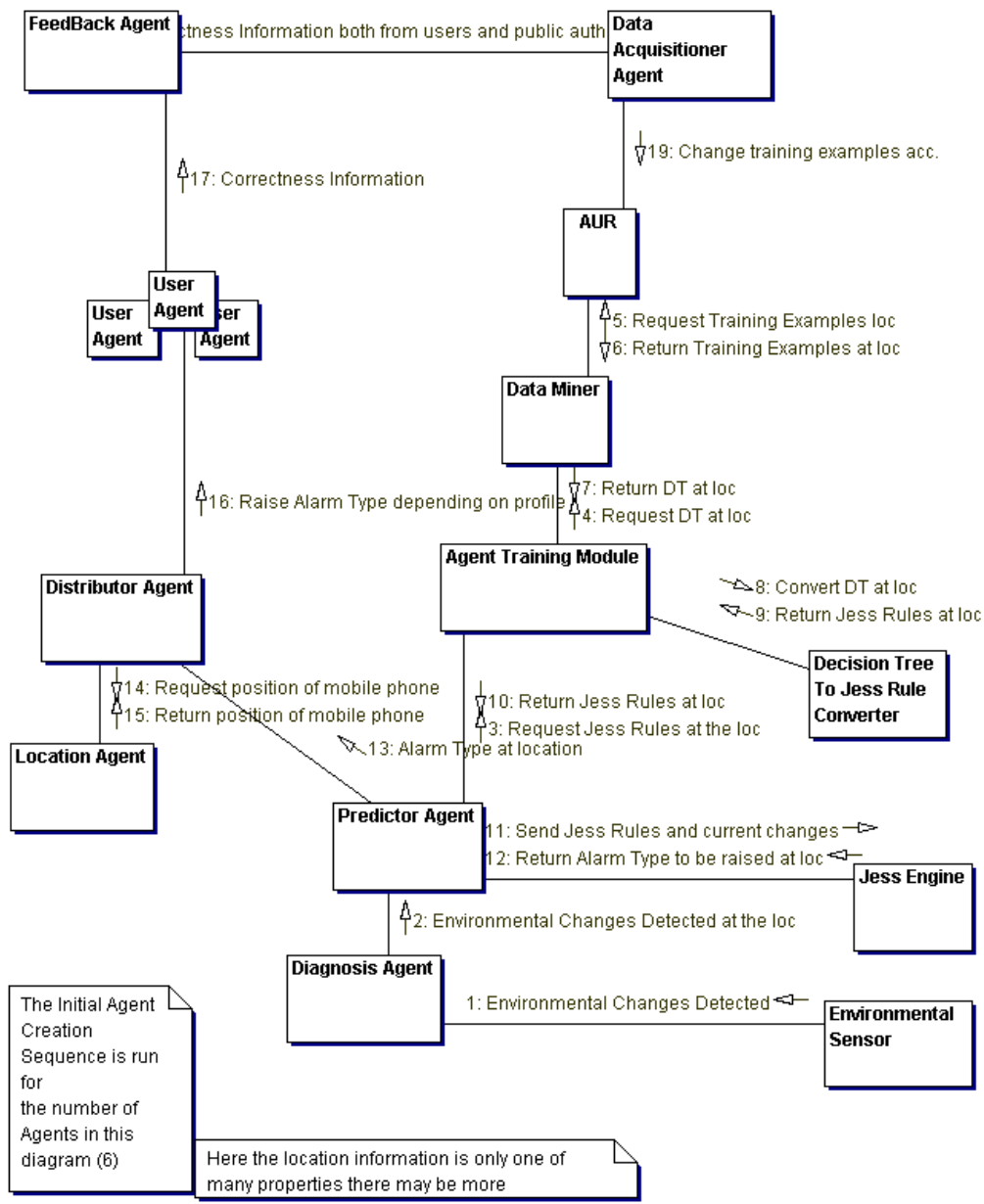

**Figure 3** Information Flow through the Agent Academy - O3RTAA Prototype

When a Predictor Agent decides it needs to update its prediction strategy, it requests a new decision structure from the Agent Training Module as a set of Java Expert System Shell (JESS) rules (3). Agent Training Module requests a prediction strategy from the Data Miner. Data Miner requests previously inserted training examples related with a certain location from the Agent Use Repository (4). This information, that is, the previously inserted training examples related with a certain location is transferred from the Agent Use Repository to the Data Miner (5,6).

Using data-mining algorithms, the Data Miner constructs a data structure and sends it to the Agent Training Module (7). The content language and the ontology are predetermined and known by both of the agents. Since the Data Miner returns the new decision strategy in a standard language, and the Predictor Agent uses a JESS rule set, it is necessary to convert one into the other. Therefore, the decision strategy obtained from the Data Miner is sent to the JESS Rule converter (8,9). The Agent Training Module forwards these rules to the Predictor Agent, which had requested them in step 3 (10).

The Predictor Agent sends these Java Expert System Shell (JESS) Rules and the current

values of environment parameters (obtained at step 2 from the Diagnosis agent and represented as Assert statements of JESS) to its JESS Engine for execution (Steps 11,12). If upon execution the expert system decides that there is a need to raise an alarm, the Predictor agent sends this alarm, along with type and location information to the Distributor Agent (13). It should be noted that the location of the change and alarm is obtained at step 2 from the Diagnosis Agent.

The User Agents hold user profile information for different subscribers. These profiles include the types of alarms the user requires, the location of the user and the delivery information and format such as HTML, CHTML, WML, or xHTML. If the user is mobile, his location information is obtained from his user Agent (steps 14,15). The User agent then, using the Ericsson Mobile Positioning System Software Development Kit, finds the location of the user as latitude and longitude. The data in this specific format is converted into a format compatible with the location information received from the Diagnosis agent provided at step 1. Based on the location of the user and the specified alarm type in the user profile, the users are alerted (step 16). The users provide feedback on the usefulness of the alarm, which are collected by the Feedback Agent to be fed into Agent Use Repository (steps 17, 18, 19). The feedback obtained from individual users affect the data in the Agent Use Repository only when it exceeds a certain threshold. The feedback coming from the public institutions, on the other hand, directly influences the Agent Use Repository.

## 6. Summary

In this paper we described the Agent Academy attempt to develop a framework through which users can create an agent community having the ability to train and retrain its own agents using Data Mining techniques. After presenting AA core components, we discussed the ontology concept and the way it affects the training procedure. We described the O3RTAA scenario, which fires environmental alarms and we realized it through the AA perspective.

We have used a multi-agent system in realizing the O3RTAA scenario because the components of the system are highly distributed, autonomous and communicate with each other through messages. Furthermore the O3RTAA system components involve decision making.

## Acknowledgements

This work has been partially supported by the European Commission IST Programme, under the contract IST-2000-31050. Authors would like to thank Mr. Miguel Alborg and his associates in IDI-EIKON, Valencia, Spain, for their assistance on the O3RTAA scenario and the Mediterranean Center for Environmental Studies Foundation (CEAM), Valencia, Spain, for the provision of the environmental data.